\documentclass[preprint,12pt]{elsarticle}
\pdfoutput=1

\newcommand{\cd}{$^{109}$Cd}

\begin{document}

\begin{frontmatter}



\title{Characterization of the Hamamatsu S8664 Avalanche Photodiode for X-Ray and VUV-light detection}


\author[ifae]{T. Lux\corref{cor1}}
\author[coimbra]{E.D.C. Freitas}
\author[coimbra]{F.D. Amaro}
\author[ifae]{O. Ballester}
\author[ifae]{G.V. Jover-Manas}
\author[ifae]{C. Mart\'in}
\author[coimbra]{C.M.B. Monteiro}
\author[ifae]{F. S\'anchez}
\author[ifae,icrea]{J.Rico}

\cortext[cor1]{Corresponding author: Thorsten.Lux@ifae.es}

\address[ifae]{Institut de Física d'Altes Energies (IFAE), 08193 Bellaterra (Barcelona), Spain}
\address[coimbra]{Centro de Instrumentra\c cao, Departamento de Física, Universidade de Coimbra, Coimbra, Portugal}
\address[icrea]{Instituci\'{o} Catalana de Recerca i Estudis Avan\c{c}ats (ICREA), 08010 Barcelona, Spain}

\begin{abstract}
We present the first operation of the Avalanche Photodiode (APD) from Hamamatsu to xenon scintillation light and to direct X-rays of 22.1 keV and 5.9 keV. A large non-linear response was observed for the direct X-ray detection. At 415 V APD bias voltage it was of about 30\% for 22.1 keV and about 45\% for 5.9 keV. The quantum efficiency for 172 nm photons has been measured to be $69 \pm 15$ \%.
\end{abstract}


\end{frontmatter}






\section{Introduction}
Avalanche photo diodes (APDs) have proven to be a good alternative to photomultiplier tubes (PMTs) in visible and VUV photon detection \cite{Lopes2001gpsc,Coelho2007444}. They are compact, consume small amounts of power and are simple to operate. APDs present also high quantum efficiency, acceptable gain, insensitivity to intense magnetic fields, resistance to high-pressure environments and low degassing properties. In particular, their low radioactivity contamination is attractive for low background experiments based on xenon (Xe), such as direct dark matter searches (XENON\cite{Aprile:2005ww}, ZEPLIN \cite{Lebedenko:2008gb}) and neutrino-less double beta decay search (EXO \cite{Neilson:2008zz}, NEXT \cite{Next:2011my}), where the radio-purity of  the photo sensors is of critical importance. \\
High Pressure TPCs based on xenon \cite{Next:2011my,Sinclair:2010zz,Nygren:2009zz} are being considered for the detection of the neutrino-less double beta decay. Gas detectors present several advantages over the liquid option.  Gaseous xenon detectors have better intrinsic energy resolution \cite{Bolotnikov1997zz} than the liquid and  the  low density media allows to track the electrons emitted from the double beta decay  reducing the background contamination from topological constraints.
Previous studies show that the operation of the detector in the so-called electroluminiscence regime allows to obtain resolutions close to the ones from the primary electron fluctuations. Electroluminiscence is achieved by accelerating the primary electrons in the xenon to an energy that produces scintillation light without entering into the charge amplification regime. This technique is well established for xenon with photomultipliers \cite{Fernandes:2010gg}  and APD \cite{Fernandes:2007rd} readouts. In this paper we evaluate the performance of the the  Hamamatsu S8664-SPL  Avalanche Photo diode sensor.  This APD is a special version of the standard product, made sensitive to xenon (172nm)  and argon (128nm) scintillating light. The APD is available in two different sizes (5x5 mm$^2$ and   10x10 mm$^2$). The small size of the sensor allows to explore the possibility of using this technology for energy measurement and tracking when laying them as an array of sensors with independent readouts \cite{Lux:2011vp}.\\
In this paper we present an independent measurement of the quantum efficiency for 172 nm photons for these devices and the first measurement of their response to direct X-rays of 22.1 keV and 5.9 keV.Although there are some applications of APDs to direct X-ray detection, e.g.\cite{Ludhova:2005}, X-ray detection with APDs was mainly investigated to measure the charge carriers produced in light measurements, using the number of charge carriers produced by the x-ray interaction in the APDs as a reference, resulting in a straight forward process to evaluate the number of charge carriers produced in the APD by the light pulse. This method has been extensively used to measure the scintillation yield in inorganic crystals \cite{Moszynski:2002a} and in noble gases \cite{Monteiro:2008}, as well as to determine the quantum efficiency of APDs \cite{Shagin:2009}.
However, non-linearities in the APD response to X-rays have to be taken into account. These are due to space charge effects, resulting from a reduction of the local electric field intensity, and from local heating, due to the point-like nature of the primary electron cloud produced by the X-ray interaction \cite{Pansart:1997,Allier:1998,Moszynski:2001}. Therefore, the non-linear response to X-rays has to be investigated for a full characterization of the present photo diodes.

\section{Experimental setup}

\begin{figure}
	\centering
		\includegraphics[width=0.7\textwidth]{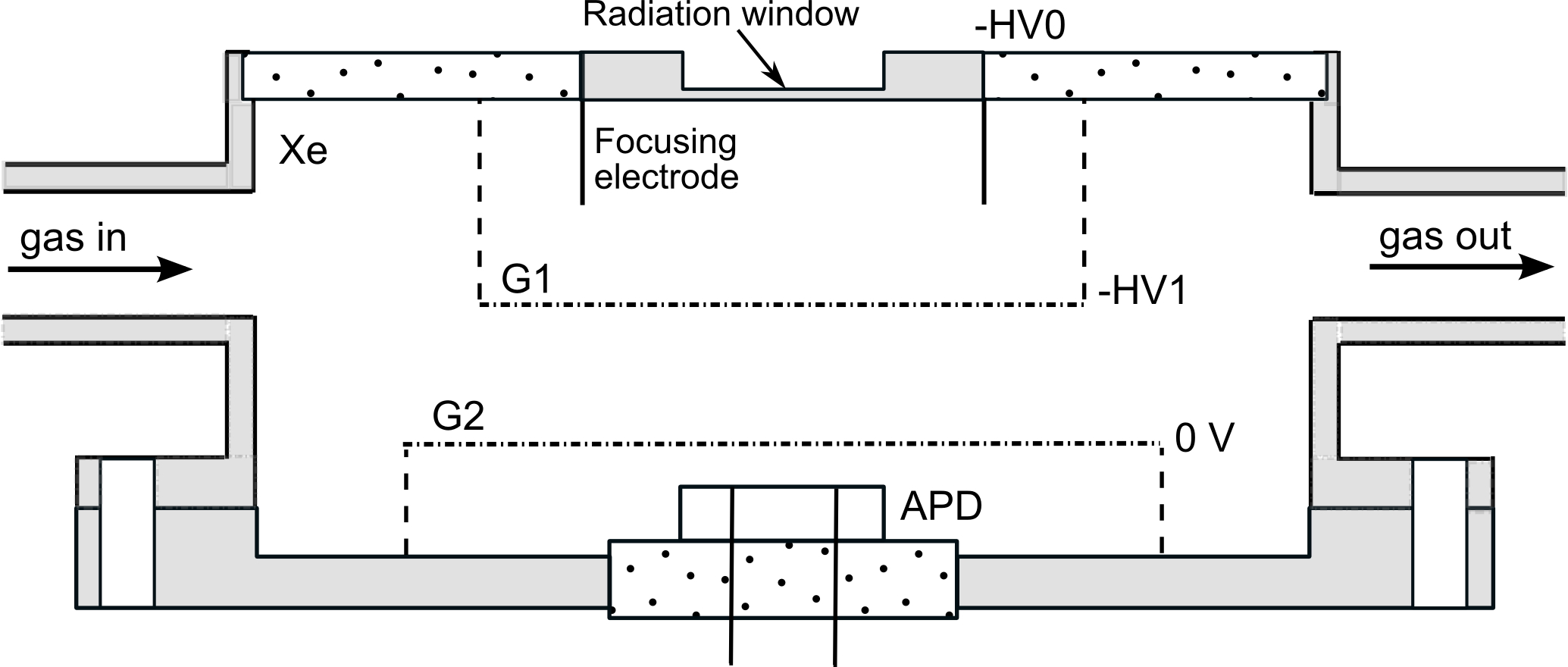}
	\caption{Schematic drawing of the GPSC used for the EL measurements.}
	\label{fig:schematic2a}
\end{figure}

Figure \ref{fig:schematic2a} depicts the schematic of the gas proportional scintillation counter (GPSC) used in this work. The detector body has a cylindrical shape of 14 cm in diameter and 5 cm in height, with a 2~mm aluminized Kapton radiation window. A stainless steel cylinder of 60 mm diameter fixes the Mesh G1, and has multiple perforations on its side surface to increase gas circulation in the drift/absorption region. The radiation window is kept at negative high-voltage HV0, while mesh G1 and its holder are kept at -HV1. Mesh G2 and detector body are grounded. Electrical insulation of the radiation window and the G1 holder is achieved using a machinable glass ceramic, Macor$^{\textregistered}$, glued to the detector body and to the window with a low vapor pressure epoxy. The voltage difference between the detector window and G1 defines the reduced electric field in the absorption/drift region, which is kept at 0.5 Vcm-1torr-1, below the xenon scintillation threshold. The scintillation region is delimited by two planar meshes: G1 and G2.  In this GPSC prototype, the absorption/ drift region and the scintillation region are 2 cm deep and 1.4 cm deep, respectively. The chamber operation parameters are shown in table \ref{Table:ChamberSettings}. X-rays interacting in the drift region produce a primary electron cloud  that drifts towards the scintillation region. Upon crossing the scintillation region, each primary electron produces, in average, a known number of scintillation photons \cite{Monteiro:2007vz}. X-ray interactions in the scintillation region will lead to scintillation pulses with lower amplitudes. These pulses result in a distortion of the Gaussian-shape pulse height distribution with a tail towards the low amplitude region. However, the peak of the pulse height distribution is not altered by this tail.
A fraction of the X-rays interact in the APD, producing a pulse height distribution that is independent of the electric fields of the GPSC, depending only on the APD biasing.

\begin{table}[htdp]
\begin{center}
\begin{tabular}{lc}
Parameter  & Value \\
\hline
Reduced electric field in drift region &   150~V/cm/bar   \\
Reduced electric field in scintillation region &  3.75~kV/cm/bar  \\
APD bias voltage &  415~V             \\
Gas pressure                &     1.07~bar      \\
\end{tabular}
\end{center}
\caption{ Description of the chamber operation parameters. }
\label{Table:ChamberSettings}
\end{table}%

\section{Method for the Q$_{eff}$ determination}
The quantum efficiency of the APD was determined by comparison of the VUV-scintillation pulse amplitudes with those resulting from direct interaction of the X-rays in the photo diodes. 
We follow here the method established in \cite{Monteiro:2007vz}.
The total number of photons produced is computed from the energy released by the {\cd}  gamma ray and the value of : 
 \begin{equation}
N_{\gamma}^{total}= N_{elec}G=\frac{E_\gamma}{w_{Xe}}G
\end{equation}
where G is the gain of the electroluminescence phase, and $N_{elec}$ is the number of primary electrons released by the Xe via ionization.  This number is obtained from the average energy needed to produce an electron-ion pair, $w$, and the total energy of the X-rays, $E_\gamma$. 
The photon yield per cm and per bar, $Y/p$, is given by the empirical formula \cite{Monteiro:2007vz}:
\begin{equation}
\frac{Y}{p}\Big[\frac{\textrm{photons}}{\textrm{cm\ bar}}\Big]=140\frac{E}{p}\Big[\frac{\textrm{kV}}{\textrm{cm\ bar}}\Big]-116\Big[\frac{\textrm{photons}}{\textrm{cm\ bar}}\Big]
\end{equation}
 
Here,  $E/p$ is the reduced electric field in the scintillation region. The gain G is then given by:
\begin{equation} \label{form_gain}
G=\frac{Y}{p}pd
\end{equation} 
Whereby $d$, the gap between the electroluminescence meshes, is given by the geometry of the detector and p is the operation pressure. For \cd, d= 1.4 cm, p=1.07 bar and a E/p=3.75 kV/cm/bar, the total number of UV photons produced is about 622000. 
The number of photons arriving to the APD, $N_{obs}$, is derived from the total number of photons emitted in 4$\pi$ and the solid angle, $\Omega_{APD}$, covered by the sensitive area of the APD. 
\begin{equation}
N_{obs}=N_{\gamma}^{total}\frac{\Omega_{APD}}{4\pi}T
\end{equation}
 
The solid angle $\Omega_{\textrm{APD}}$ and the transparency $T$ of the electroluminescence mesh are estimated from a Monte Carlo program described in section \ref{sec_mc}.
The quantum efficiency Qeff, defined as the number of free electrons produced in the APD, $N_{e,sci}$,  per VUV photon, is then given by:
\begin{eqnarray}
Q_{eff}&=&\frac{N_{e,sci}}{N_{obs}} \\
       &=&\frac{A_{UV}}{A_{XR}}\frac{N_{XR}}{N_{obs}}
\label{QEEq}
\end{eqnarray}  
Here, $A_{UV}$ and $A_{XR}$ are the peak position of the UV and the direct X-ray peak in the pulse-height spectrum and $N_{XR}$ is the number of electrons released in the silicon by the direct absorption of a X-ray.  This number can be calculated from the X-ray energy, $E_{XR}$ (22.1 keV for \cd), the energy needed to produce a electron-ion pair in silicone, $w_{Si}$ (3.62 eV \cite{Knoll:2000}) and a quenching factor $Q_f$. 
\begin{equation}
N_{XR}=\frac{E_{XR}}{w_{Si}}Q_f
\end{equation} 
The latter takes into account high charge density effects as they can occur in the absorption of X-rays. This factor depends on the type of APD, the applied bias voltage and the X-ray energy and has to be determined experimentally as described in the following section.

\section{Non-linear response to X-rays}
\begin{figure}
	\centering
		\includegraphics[width=0.8\textwidth]{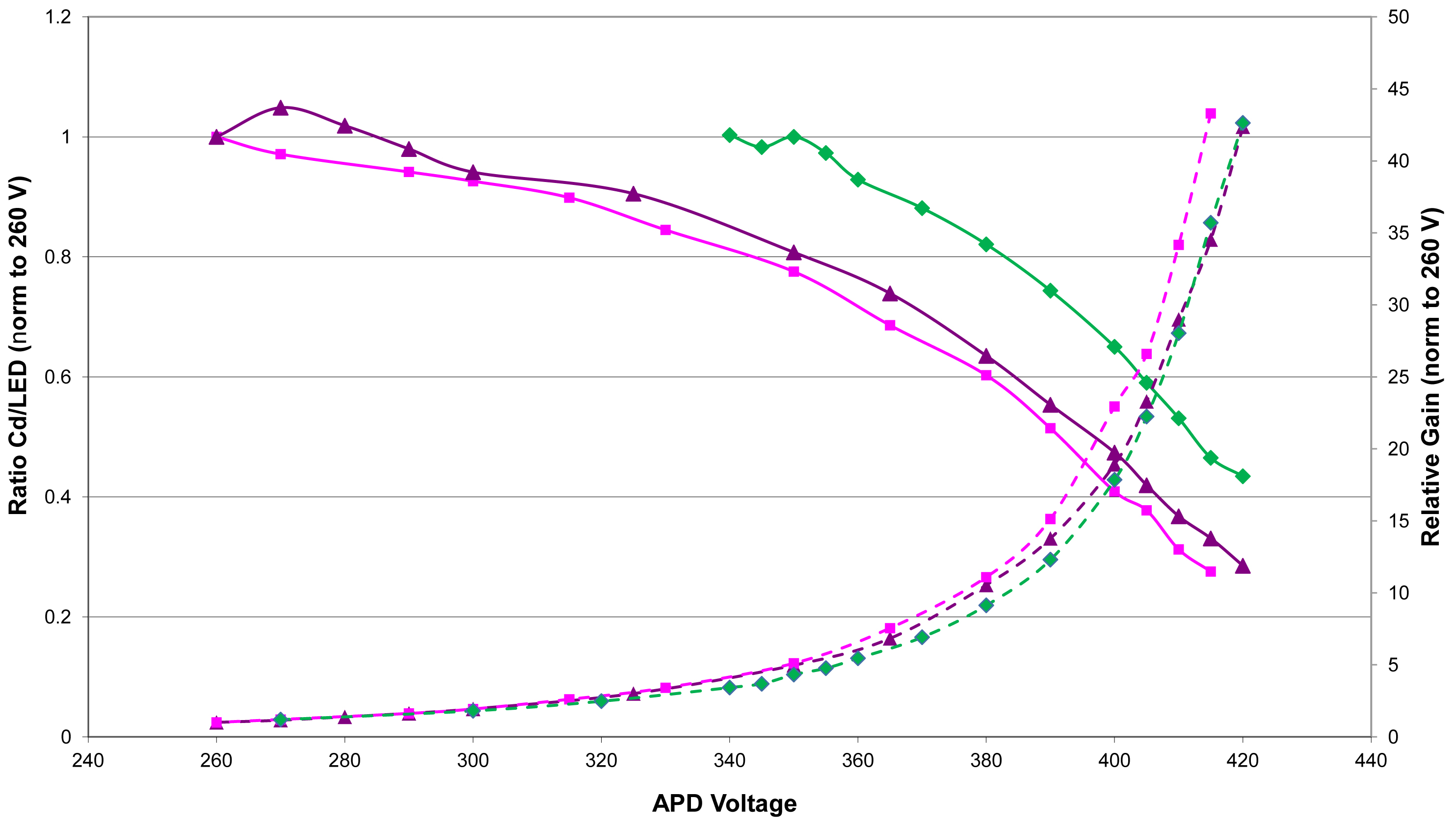}
	\caption{Ratio of X-ray peak position and LED peak position as function of the APD bias voltage (solid line) and the LED light gain (dashed line). The light gain is normalized to the value at 260 V. The purple and pink lines show the results for 22.1 keV from a $^{109}$Cd source for two different APDs, normalized to the value at 260 V, while the green line shows the result for 5.9 keV, normalized to the value at 350 V.}
	\label{fig:cd_quench_vs_apdvolt}
\end{figure}

\begin{figure}
	\centering
		\includegraphics[width=0.8\textwidth]{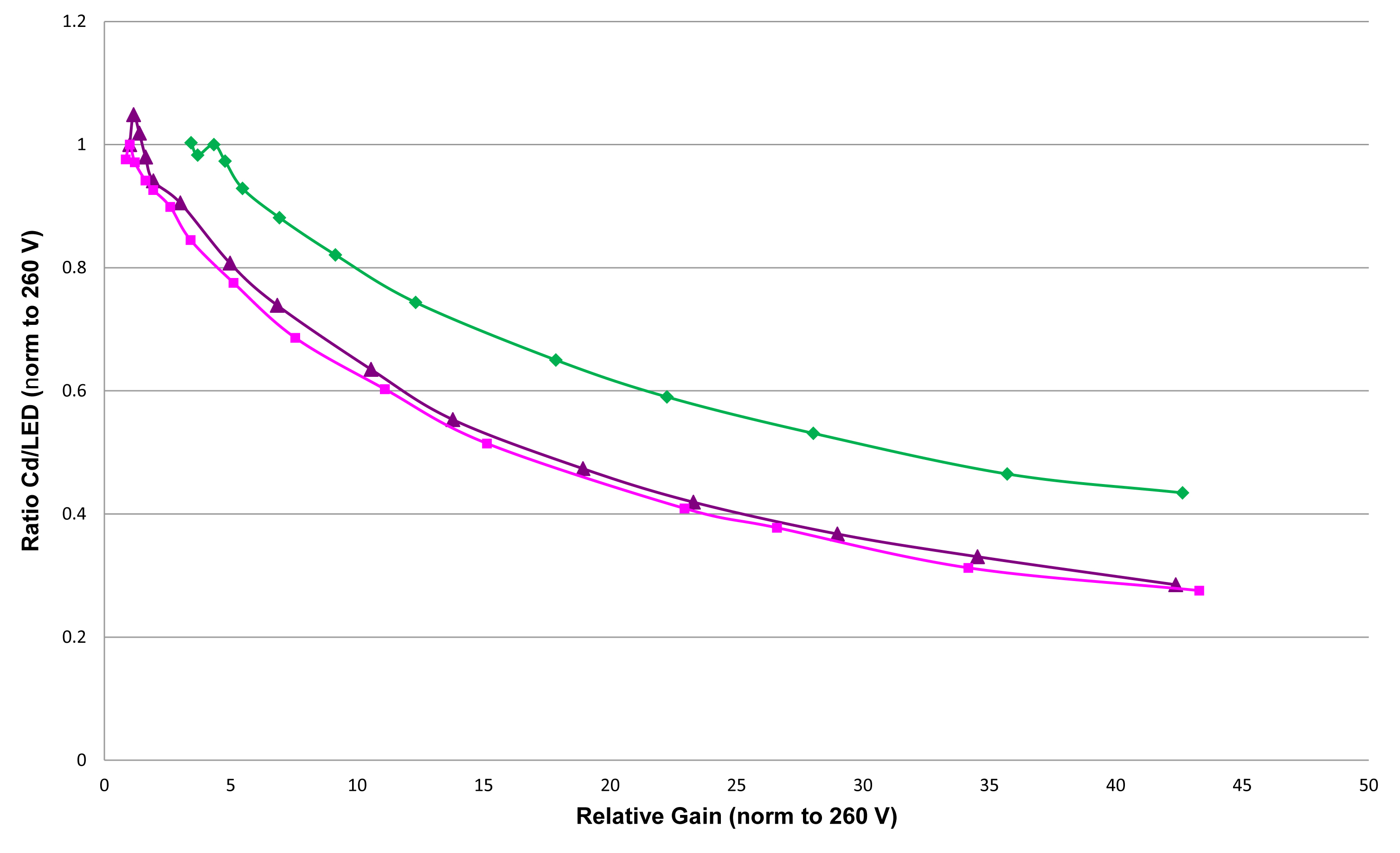}
	\caption{The quenching factor as function of LED light gain, normalized to the value at 260 V. As before The purple and pink lines show the results for 22.1 keV from a $^{109}$Cd source for two different APDs, normalized to the value at 260 V, while the green line shows the result for 5.9 keV, normalized to the value at 350 V.}
	\label{fig:cd_quench_vs_gain}
\end{figure}

The non-linear response to X-rays was measured for the first time for this type of APDs. 
The method used follows the one described in \cite{Mosznski:2002}. The APD was mounted in a gas tight box and flushed with dry gas. In the box also a radioactive source, $^{109}$Cd or $^{55}$Fe, and a green LED of about 520 nm were installed in such a way that the APD could be illuminated simultaneously from both. The APD signal was processed with a charge sensitive ORTEC 142B preamplifier, an ORTEC 673 amplifier and the spectrum was recorded with a 8001A multichannel-analyzer from AMPTEK. 
The non-linear response was determined by simultaneous monitoring the amplitudes due to the interactions in the APD of the LED light pulses as well as the X-rays. From the development of the ratio of the position of the LED and the X-ray peaks as function of the applied voltage the non-linear response was determined. We call this ratio the quenching factor.
During the measurements the temperature was stable within 2$^{\circ}\ C$. Fig.~\ref{fig:cd_quench_vs_apdvolt} shows the X-ray quenching factor and the LED light gain, normalized to the value at 260 V, as function of the APD bias voltage.
The differences between the two APDs are vanishing when one plots the quenching factor as function of the gain (fig.~\ref{fig:cd_quench_vs_gain}). This behavior is expected since the quenching is caused by local high charge densities in the APD and it should be independent of the bias voltage applied to achieve this gain. 
The result of this study is that the quenching factor for this type of APD is $0.29 \pm 0.04$ for 22.1 keV and $0.46 \pm 0.07$ for 5.9 keV X-rays at 415 V. These values are significantly larger than the ones obtained with APDs from Advanced Photonics Inc. (API) \cite{Fernandes:2007rd, Moszynski:2000a} indicating that these devices are less suitable for direct X-ray detection.  

\section{Monte Carlo simulation of the photon acceptance} \label{sec_mc}

We have developed a Monte Carlo simulation program describing all the key elements of our experimental setup. It simulates both primary electron production and its drift and scintillation light production, light transport and the detection by the APD. The X-ray profile for the $^{109}$Cd is simulated considering the acceptance limitation by the collimator and the exponential absorption in the gas.  Primary electrons are transported in the gas following a Gaussian transverse diffusion model. The scintillation production is simulated photon-by-photon along the transport of the primary electrons within the EL gap. The light emission is  isotropic and the total number of photons produced is calculated via formula \ref{form_gain}, the absorption of the mesh between the EL gap and the APDs is also accurately simulated. \\ 
  The code integrates the total number of photons arriving to the APD under the assumption that the collection area is 5x5~mm$^2$. The photon acceptance is then obtained as the ratio between this number and the total number of photons produced in the EL gap.\\
  The input values to the simulation are shown in Table \ref{Table:MC} with the precision we have determined them. The value of the APD acceptance is estimated to be  $ 0.0123 \pm 0.0014  (syst)$, the systematic error is estimating varying the different parameters according their known accuracy as shown in Table 2.

\begin{table}[htdp]
\begin{center}
\begin{tabular}{ l c }
Parameter   & value     \\
\hline 
Drift volume distance  &   ($19.9\pm0.2$) mm       \\
EL mesh gap length   &    ($14.0\pm0.2$) mm       \\
Distance between lower mesh and APD &  ($4.7\pm0.5$) mm  \\
Mesh wire radius   &     ($0.04\pm0.004$) mm       \\
Mesh wire pitch   &     ($0.90\pm0.09$) mm       \\
Transverse diffusion  &      $2346\ \mu\textrm{m}/\sqrt{cm}$ \cite{biagi:1999}  \\
\end{tabular}
\end{center}
\caption{ Simulation parameter with its defined value and estimated precision.}
\label{Table:MC}
\end{table}%

\section{Results and discussion} 

The final value of the Hamamatsu S8664-SPL quantum efficiency is derived from equation \ref{QEEq}. The values for the different elements in the equation are compiled in table  \ref{Table:Values}. The result for the quantum efficiency is $69 \pm 15$ \% for 172 nm photons using a $^{109}$Cd source, which is an independent confirmation of the value of about 80\% quoted by Hamamatsu. 
This high quantum efficiency makes this device very attractive for the light readout of xenon-based detectors. Further tests with an array of 5 APDs and measurements in argon are currently being performed to investigate the full potential of this sensor.\\   
We have also studied for the first time the non-linear response of this device to X-rays. The quenching factor shows a large non-linear response to direct X-rays. 
Relative light to X-ray  gain reduction to 45\% and 30\%  were measured for 5.9 and 22.1 keV for relative gains of around 40. These values are much higher than those obtained for other type of APDs, denoting a much higher presence of non-linearity effects in X-ray detection with this type of APDs.
  
\begin{table}[htdp]
\begin{center}
\begin{tabular}{lc}
Parameter & Value \\
\hline
A$_{UV}$                            &  $145.5 \pm 1.5 $ \\
A$_{XR}$                            &  $48.1 \pm 3.4 $ \\
N$_{XR}$                            &  $6105$ \\
Number of photons           &  $622000 \pm 62200 $ \\
Acceptance                        &  $0.0123 \pm 0.0014$ \\
Quenching factor (415 V/ 22.1 keV)              &  $0.29 \pm 0.04  $ \\
\end{tabular}
\end{center}
\caption{ Values of the parameters used for the quantum efficiency calculation. }
\label{Table:Values}
\end{table}%

\section*{Acknowledgments}
The authors acknowledge the support received by the CONSOLIDER INGENIO Project CSD2008-0037 (CUP), from MICINN (Spain) through project FPA2009-13697-C04-03, from FCT (Portugal) and FEDER under COMPETE program, through project PTDC/FIS/103860/2008.

\bibliographystyle{model1a-num-names}
\bibliography{all}

\end{document}